\documentstyle[12pt]{article}

\begin{document}

\newcommand{\nc}[2]{\newcommand{#1}{#2}}
\newcommand{\ncx}[3]{\newcommand{#1}[#2]{#3}}
\ncx{\pr}{1}{#1^{\prime}}
\nc{\nl}{\newline}
\nc{\np}{\newpage}
\nc{\nit}{\noindent}
\nc{\be}{\begin{equation}}
\nc{\ee}{\end{equation}}
\nc{\ba}{\begin{array}}
\nc{\ea}{\end{array}}
\nc{\bea}{\begin{eqnarray}}
\nc{\eea}{\end{eqnarray}}
\nc{\nb}{\nonumber}
\nc{\dsp}{\displaystyle}
\nc{\bit}{\bibitem}
\nc{\ct}{\cite}
\ncx{\dd}{2}{\frac{\partial #1}{\partial #2}}
\nc{\pl}{\partial}
\nc{\dg}{\dagger}
\nc{\cL}{{\cal L}}
\nc{\cD}{{\cal D}}
\nc{\cF}{{\cal F}}
\nc{\cG}{{\cal G}}
\nc{\cJ}{{\cal J}}
\nc{\cQ}{{\cal Q}}
\nc{\tB}{\tilde{B}}
\nc{\tD}{\tilde{D}}
\nc{\tH}{\tilde{H}}
\nc{\tR}{\tilde{R}}
\nc{\tZ}{\tilde{Z}}
\nc{\tg}{\tilde{g}}
\nc{\tog}{\tilde{\og}}
\nc{\tGam}{\tilde{\Gam}}
\nc{\tPi}{\tilde{\Pi}}
\nc{\tcD}{\tilde{\cD}}
\nc{\tcQ}{\tilde{\cQ}}
\nc{\ag}{\alpha}
\nc{\bg}{\beta}
\nc{\gam}{\gamma}
\nc{\Gam}{\Gamma}
\nc{\bgm}{\bar{\gam}}
\nc{\del}{\delta}
\nc{\Del}{\Delta}
\nc{\eps}{\epsilon}
\nc{\ve}{\varepsilon}
\nc{\zg}{\zeta}
\nc{\th}{\theta}
\nc{\vt}{\vartheta}
\nc{\Th}{\Theta}
\nc{\kg}{\kappa}
\nc{\lb}{\lambda}
\nc{\Lb}{\Lambda}
\nc{\ps}{\psi}
\nc{\Ps}{\Psi}
\nc{\sg}{\sigma}
\nc{\spr}{\pr{\sg}}
\nc{\Sg}{\Sigma}
\nc{\rg}{\rho}
\nc{\fg}{\phi}
\nc{\Fg}{\Phi}
\nc{\vf}{\varphi}
\nc{\og}{\omega}
\nc{\Og}{\Omega}
\nc{\Kq}{\mbox{$K(\vec{q},t|\pr{\vec{q}\,},\pr{t})$ }}
\nc{\Kp}{\mbox{$K(\vec{q},t|\pr{\vec{p}\,},\pr{t})$ }}
\nc{\vq}{\mbox{$\vec{q}$}}
\nc{\qp}{\mbox{$\pr{\vec{q}\,}$}}
\nc{\vp}{\mbox{$\vec{p}$}}
\nc{\va}{\mbox{$\vec{a}$}}
\nc{\vb}{\mbox{$\vec{b}$}}
\nc{\Ztwo}{\mbox{\sf Z}_{2}}
\nc{\Tr}{\mbox{Tr}}
\nc{\lh}{\left(}
\nc{\rh}{\right)}
\nc{\ld}{\left.}
\nc{\rd}{\right.}
\nc{\nil}{\emptyset}
\nc{\bor}{\overline}
\nc{\ha}{\hat{a}}
\nc{\da}{\hat{a}^{\dg}}
\nc{\hb}{\hat{b}}
\nc{\db}{\hat{b}^{\dg}}
\nc{\hN}{\hat{N}}
\ncx{\abs}{1}{\left| #1 \right|}

\pagestyle{empty}

\begin{flushright}
NIKHEF-H/93-04\\
DAMTP R92/43 \\
\end{flushright}

\begin{center}
{\LARGE {\bf SUSY in the sky}} \\

\vspace{3ex}

{\large G.W.\ Gibbons\footnote{e-mail: gwg1@amtp.cam.ac.uk}}  \\
        DAMTP, Cambridge U.K. \\

\vspace{2ex}

{\large R.H.\ Rietdijk\footnote{e-mail: r.h.rietdijk@durham.ac.uk}} \\
        Physics Dept., Univ.\ of Durham U.K. \\

\vspace{2ex}

{\large J.W.\ van Holten\footnote{e-mail: t32@nikhef.nl}} \\
        NIKHEF-H, Amsterdam NL \\

\vspace{5ex}

{\bf Abstract} \\
\end{center}

\nit
{\small
Spinning particles in curved space-time can have fermionic symmetries
generated by the square root of bosonic constants of motion other than the
Hamiltonian. We present a general analysis of the conditions under which such
new supersymmetries appear, and discuss the Poisson-Dirac algebra of the
resulting set of charges, including the conditions of closure of the new
algebra. An example of a new non-trivial supersymmetry is found in black-hole
solutions of the Kerr-Newman type and corresponds to the Killing-Yano tensor,
which plays an important role in solving the Dirac equation in these black-hole
metrics. }

\np

\pagestyle{plain}
\pagenumbering{arabic}

\section{Introduction}{\label{S.1}}

In this paper we investigate the symmetries of classical space-times in terms
of
the motion of pseudo-classical spinning point particles in a curved Lorentzian
manifold. Pseudo-classical spinning point particles are described by the $d =
1$
supersymmetric extension of the simple (spinless) relativistic point particle,
as developed in \ct{BM}-\ct{vH}. The general relations between space-time
symmetries and the motion of spinning point particles has been analysed in
detail in \ct{RH1}-\ct{RH3}. These methods may be applied to any space-time,
but
since undoubtedly the most important solution of the Einstein vacuum equations
in 4 dimensions is the Kerr solution, which represents the gravitational field
of an isolated rotating black hole, the detailed applications considered in the
later sections of this paper are taylored more in particular to that case and,
almost equally interesting from the theoretical point of view, to the
Kerr-Newman solutions of the combined Einstein-Maxwell equations. With the
exception of the charged multi-black-hole metrics found by Papapetrou and
Majumdar, these are believed to be the unique stationary and asymptotically
flat
solutions of the Einstein-Maxwell equations which are regular outside the
(equally regular) event horizon. If the cosmic censorship hypothesis holds,
then
it seems likely that these metrics represent the only possible final exterior
equilibrium states of gravitational collapse.

The Kerr-Newman solutions are invariant under two continuous symmetries: time
translations and rotations about an axis of symmetry, which are generated by
Killing fields $K^{\mu}$ and $M^{\mu}$, respectively. These symmetries give
rise to two constants of motion: energy $E$ and angular momentum $J$, for
particles orbiting in these backgrounds. Both constants of motion are
linear in the particle's 4-momentum $p_{\mu}$:

\be
E\, =\, - K^{\mu} p_{\mu},
\label{1.1}
\ee

\nit
and

\be
J\, =\, M^{\mu} p_{\mu}.
\label{1.2}
\ee

\nit
It came, therefore, as a considerable surprise when Carter succeeded in showing
that because of the existence of a further constant of motion, quadratic in the
4-momentum, the equations for the geodesics and the orbits of charged particles
constituted a completely integrable system in the sense of Liouville
\ct{Carter}. Carter's constant of motion has the form

\be
Z\, =\, \frac{1}{2}\, K^{\mu\nu}\, p_{\mu} p_{\nu},
\label{1.3}
\ee

\nit
which commutes with the covariant Hamiltonian

\be
H\, =\, \frac{1}{2}\, g^{\mu\nu} p_{\mu} p_{\nu}
\label{1.4}
\ee

\nit
in the sense of Poisson brackets. This is guaranteed because $K^{\mu\nu}$ is a
symmetric second-rank contravariant tensor field satisfying a generalised
Killing equation \ct{RH1,RH2}, a class known as Stackel-Killing tensors. The
four constants of motion $(E, J, Z, H)$ now form a mutually Poisson-commuting
set of functions on the cotangent bundle, two of which are linear and two of
which are quadratic in the momenta.

Historically, it was shown by Carter \ct{Carter2} that the Klein-Gordon
equation
with minimal coupling to the electromagnetic field is soluble by separation of
variables in these backgrounds. Also various other field equations were shown
to
separate. Most significantly for our present purpose, Chandrasekhar achieved
separation of the Dirac-equation \ct{Ch}.

Carter was able to reinterpret the separability of the Klein-Gordon equation
for
charged particles in terms of a second order differential operator constructed
from the Stackel-Killing tensor $K^{\mu\nu}$ similar to the construction of the
Laplace-Beltrami operator from the contravariant components of the metric
$g^{\mu\nu}$: one considers the operator $\hat{K}$ operating on a scalar wave
function $\Ps$ by

\be
\hat{K}\, \Ps\, =\, D_{\mu} K^{\mu\nu} D_{\nu} \Ps.
\label{1.5}
\ee

\nit
Here $D_{\mu}$ is a covariant derivative including minimal coupling to the
electromagnetic field. This operator $\hat{K}$ now commutes with the covariant
Laplacian

\be
\Del\, =\, D_{\mu} g^{\mu\nu} D_{\nu}.
\label{1.6}
\ee

\nit
It also commutes with the Lie derivatives with respect to the Killing vector
fields $(K^{\mu}, M^{\mu})$, acting on a scalar wave function by linear
differential operators:

\be
K^{\mu} \pl_{\mu}\, \Ps, \hspace{3em} M^{\mu} \pl_{\mu}\, \Ps.
\label{1.7}
\ee

\nit
Thus Carter established a quantum version of his earlier classical results.

In subsequent work with McLenaghan \ct{CMcL}, Carter constructed a linear
differential operator which {\em commutes} with the Dirac operator in the
Kerr-Newman background. The construction of this operator depended upon another
remarkable fact, discovered by Penrose and Floyd, that the Stackel-Killing
tensor $K^{\mu\nu}$ has a certain square root, which defines a Killing-Yano
2-form $f_{\mu\nu} = - f_{\nu\mu}$ such that

\be
K^{\mu}_{\:\:\nu}\, =\, f^{\mu}_{\:\:\lb} f^{\lb}_{\:\:\nu}.
\label{1.8}
\ee

\nit
Here indices are raised and lowered with the space-time metric $g_{\mu\nu}$
and its inverse.

These remarkable discoveries were at the same time both useful and mysterious.
On the one hand they made possible a whole range of calculations, both
classical
and quantum mechanical, which can and are being applied to various physical
processes near black holes in our universe. On the other hand they raise a
number of theoretical questions, including what is the {\em classical} counter
part of Carter and McLenaghan's work on the Dirac equation and what, if any, is
the relation between {\em supersymmetry} and the mysterious Killing-Yano type
square root of the Stackel-Killing tensor. It is the purpose of this paper to
address these two questions and to show how supersymmetric particle mechanics
involving classically anti-commuting Grassmann variables can be used to
understand this aspect of black holes. Furthermore it is shown how this
construction fits into the more general framework for finding constants of
motion which involve higher-rank tensors and p-forms, as outlined in
refs.\ct{RH1,RH2}. The passage to quantum mechanics is then made as
discussed for example in \ct{BM}-\ct{vH}, and no special problems are expected
to occur.

The conclusion of our analysis is, that the existence of the Killing-Yano
tensor
discovered by Penrose and Floyd and its properties may be understood in a {\em
systematic} way as a particularly interesting example of a more general
phenomenon, the appearance of an additional supersymmetry in the usual $N = 1$
supersymmetric extension of point particle mechanics in curved space-time. It
was discovered by Zumino \ct{Z}, that demanding the system to admit an
additional supersymmetry of the usual type, generating an $N = 2$
super-Poincar\'{e} algebra, restricts the background metric (which he took to
have Euclidean signature) to correspond to a K\"{a}hler manifold. This implies
that the metric $g_{\mu\nu}$ should admit a covariantly constant 2-form
$f_{\mu\nu} = - f_{\nu\mu}$:

\be
D_{\lb}\, f_{\mu\nu}\, =\, 0.
\label{1.9}
\ee

\nit
Quite aside from the problem of the space-time signature (which could possibly
be changed to Kleinian, $++--$) this condition is far too restrictive to be
useful in any direct astrophysical application. However, it is possible to make
the weaker demand that an extra supersymmetry exists but not to prejudge the
algebra it is to satisfy. This weaker condition {\em is} compatible with a
Lorentzian signature and gives rise to the correspondingly weaker condition

\be
D_{\mu}\, f_{\nu\lb}\, +\, D_{\nu}\, f_{\mu\lb}\, =\, 0,
\label{1.10}
\ee

\nit
where the covariant tensor field $f_{\mu\nu}$ need not necessarily be
anti-symmetric. In the case that $f_{\mu\nu}$ is anti-symmetric, we are led
precisely to the condition for the existence of a second-rank Killing-Yano
tensor of the type found by Penrose and Floyd, and exploited so succesfully
by Carter and McLenaghan. Moreover it is easy to see that the expression
given by them for a first-order differential operator which {\em commutes} with
the Dirac operator $\gam^{\mu} D_{\mu}$, to wit:

\be
i \gam_{5} \gam^{\mu} \lh f_{\mu}^{\:\:\nu} D_{\nu}\, -\, \frac{1}{6}\,
  \gam^{\nu} \gam^{\lb} D_{\lb} f_{\mu\nu} \rh,
\label{1.11}
\ee

\nit
{\em anti-commutes} with the Dirac operator after multiplication by $\gam_{5}$,
which is more natural from the point of view of supersymmetry.

As is well-known, in the quantum theory of the spinning point particle the
first
Poincar\'{e}-type supercharge is represented by a multiple of $\gam_{5}$ times
the Dirac operator acting on spinor fields on the space-time manifold. In the
special case one is dealing with a K\"{a}hler metric, for which $D_{\lb}
f_{\mu\nu} = 0$, the second Poincar\'{e} supercharge is then given by a
multiple
of

\be
i \gam_{5} \gam^{\nu}\, f^{\:\:\mu}_{\nu}\, D_{\mu}.
\label{1.12}
\ee

\nit
This operator is constructed from the Levi-Civita connection on spinor fields
in
terms of the gamma matrices $i f^{\:\:\mu}_{\nu} \gam^{\nu}$, which have been
rotated with respect to those appearing in the Dirac operator using a complex
structure $f^{\mu}_{\:\:\nu}$, provided it is normalised to satisfy

\be
f^{\mu}_{\:\:\lb} f^{\lb}_{\:\:\nu}\, =\, - \del^{\mu}_{\nu}.
\label{1.13}
\ee

\nit
The expression of Carter and McLenaghan is a natural generalisation of eq.\
(\ref{1.12}) and is the quantum version of the generalised second supersymmetry
we will show to exist in the Kerr-Newman background.

At this point we should remark, that space-time supersymmetry has previously
been applied to charged black holes in the context of $N = 2$ supergravity
theory. The application of world-line supersymmetry in this paper seems at
first sight to be unrelated to that work. For example, our results concerning a
`hidden' supersymmetry related to the motion of spinning point particles are
applicable to {\em all} members of the Kerr-Newman family of black-hole
solutions, while the Killing spinors giving rise to symmetries of the solutions
of supergravity field equations arise only in the case of extreme solutions (or
indeed naked singularities) whose mass and charge in suitable units are equal.

It seems therefore that our scheme enables us for the first time to succesfully
apply supersymmetry to a problem of genuine astrophysical interest. Indeed it
is
widely believed that our universe contains large numbers of macroscopic
rotating
black holes with masses up to perhaps $10^{6}$ solar masses, all described to
good approximation by the Kerr metric. Hence the title of our paper. Of course
we do not claim this supersymmetry to be connected with any hypothetical
supersymmetry acting at elementary particle scales.

Another remark we should like to make is that from the point of view of
Hamiltonian mechanics the two symmetric contravariant tensor fields
$g^{\mu\nu}$
and $K^{\mu\nu}$ appear on a completely symmetrical footing. For example, the
Poisson commutativity of their associated constants of motion may be expressed
on the space-time manifold entirely in terms of the vanishing of their Schouten
bracket, which is an (anti-)symmetrical expression in terms of their partial
derivatives not containing any covariant (i.e.\ Levi-Civita) derivatives: it is
covariant as it stands. One might therefore be tempted to conjecture, at least
for the Kerr solution, that this symmetry might run even deeper. This however
does not seem to be the case, at least for the Kerr solution, because quite
independently of any global questions, the two tensors differ in their purely
local properties. Consider the symmetric covariant tensor fields given by their
inverses. In the case of $g_{\mu\nu}$ it satisfies the vacuum Einstein
equations. In the case of $(K^{-1})_{\mu\nu}$, substitution of this symmetric
tensor as a metric into the Einstein tensor does {\em not} lead to a solution
of the vacuum Einstein equations.  In fact one may also compute the Ricci
tensor
of $g_{\mu\kg} g_{\nu\lb} K^{\kg\lb}$. It does not vanish either. Nevertheless,
the study of these `linked geometries' is an interesting one which will be the
subject of a separate publication.

The plan of this paper is as follows. In sect.\ \ref{S.2} we review the
formalism of pseudo-classical spinning point particles in an arbitrary
background space-time, using anticommuting Grassmann variables to describe the
spin degrees of freedom. We should point out here that the equations of motion
apply in practice as a suitable semi-classical approximation to the dynamics of
a massive spin-1/2 particle such as the electron. The question of what is the
correct equation of motion for an extended macroscopic object with angular
momentum, such as a spinning neutron star, is a separate one not addressed in
this paper, although the qualitative features will surely be similar. In
particular, the Killing tensors of Stackel and Yano type are also useful in
dealing with such macroscopic spinning bodies. Another related question is what
is the correct equation of motion for a classical massless spin-1 particle such
as a photon. It is well-known that one may take a suitable WKB-approximation to
the classical Maxwell equations and discover that the appropriate equation for
the classical rays is that they move along null geodesics, and that the
polarisation vector is parallel transported along the null geodesic. Again both
types of Killing tensors are useful to deal with these equations and have been
used for that purpose. For example, the rotation of the plane of polarisation
of
radio waves passing near a spinning black hole has been calculated using the
Penrose-Floyd Yano-Killing tensor \ct{ITT,CPS}. Thus our results are also
relevant to that case though the equations of motion themselves are not the
same.

In sect.\ \ref{S.3} we review the general relation between symmetries,
supersymmetries and constants of motion for these equations.

In sec.\ \ref{S.4} we take up the question of the existence of extra
supersymmetries and their algebras. Supersymmetries are shown to depend on
the existence of a second rank tensor field $f_{\mu\nu}$ which we refer to
as {\em f-symbols}.

The general properties of {\em f}-symbols are investigated in sect.\ \ref{S.5}
and their relation to Killing-Yano tensors is pointed out. The results of this
section are rather general. They may, for instance, readily be applied to the
special case of hyper-K\"{a}hler four-manifolds, though we shall not do so in
this paper. Hyper-K\"{a}hler manifolds admit three covariantly constant 2-forms
and therefore exhibit $N = 4$ supersymmetry. This would be relevant to
Kaluza-Klein monopoles, described by the Taub-NUT metrics, or the moduli space
of two BPS Yang-Mills monopoles, as given by the Atiyah-Hitchin metric.

Finally, in sect.\ \ref{S.8} we turn to the main (astrophysically relevant)
application of this paper and exhibit the exact form of the constants of motion
in the Kerr-Newman geometry.

\section{Spinning Particles in Curved Space-Time}{\label{S.2}}

As is well-known the pseudo-classical limit of the Dirac theory of a spin-1/2
fermion in curved space-time is described by the supersymmetric extension of
the
ordinary relativistic point-particle \ct{BM}-\ct{vH}. The configuration space
of this theory is spanned by the real position variables $x^{\mu}(\tau)$ and
the
Grassmann-valued spin variables $\ps^{a}(\tau)$, where $\mu$ and $a$ both run
from $1,...,d$, with $d$ the dimension of space-time. The index $\mu$ labels
the
space-time co-ordinates and components of vectors in space-time (world
vectors),
and $a$ labels the components of vectors in tangent space (local Lorentz
vectors), among them the anti-commuting spin-co-ordinates. These types of
vectors
can be converted into each other by the vielbein $e_{\mu}^{\:\:a}(x)$ and its
inverse $e_{\:\:a}^{\mu}(x)$; for example it is sometimes convenient to
introduce the object

\be
\ps^{\mu}(x)\, =\, e_{\:\:a}^{\mu}(x)\, \ps^{a},
\label{2.0}
\ee

\nit
transforming under general co-ordinate and local Lorentz transformations as
a world vector rather than a local Lorentz vector. The world-line parameter
$\tau$ is the invariant proper time,

\be
c^{2} d\tau^{2}\, =\, - g_{\mu\nu}(x)\, d x^{\mu} d x^{\nu},
\label{2.1}
\ee

\nit
in our conventions; by choosing units such that $c = 1$ this constant no longer
explicitly appears in our equations.

The equations of motion of the pseudo-classical Dirac particle can be derived
from the Lagrangian

\be
L = \frac{1}{2}\, g_{\mu\nu}\, \dot{x}^{\mu} \dot{x}^{\nu} + \frac{i}{2}\,
  \eta_{ab}\, \ps^{a} \frac{D\ps^{b}}{D\tau},
\label{2.2}
\ee

\nit
where $\eta_{ab}$ is the flat-space (Minkowski) metric. Here and in the
following the overdot denotes an ordinary proper-time derivative $d/d\tau$,
whilst the covariant derivative of the spin variable, transforming as a local
Lorentz vector, is

\be
\frac{D\ps^{a}}{D\tau}\, =\, \dot{\ps}^{a}\, -\, \dot{x}^{\mu}\,
                             \og_{\mu\:\:b}^{\:\:a}\, \ps^{b},
\label{2.3}
\ee

\nit
with $\og_{\mu\:\:b}^{\:\:a}$ the spin connection. In order to fix the dynamics
completely one has to add the condition expressed by eq.(\ref{2.1}), which is
equivalent to the mass-shell condition, plus others necessary to select the
physical solutions of the equations of motion. For example, the restriction
that
spin be space-like reads

\be
{\cal Q}\, \equiv\, e_{\mu a}\, \dot{x}^{\mu}\, \ps^{a}\, =\, 0,
\label{2.3.1}
\ee

\nit
implying that $\ps$ has no time-component in the rest frame. These
supplementary
conditions have to be compatible with the equations of motion derived from the
Lagrangian $L$ \ct{RH2,RH3}; however, in our formulation of spinning particle
dynamics they are only to be imposed {\em after} solving these
equations\footnote{Actually, we work in a gauge-fixed formulation of the
super-reparametrization invariant theory, hence we have to impose the gauge
conditions as separate restrictions on the dynamics.}.

The configuration space of spinning particles spanned by $(x^{\mu}, \ps^{a})$
is sometimes refered to as spinning space. The solutions of the Euler-Lagrange
equations derived from the Lagrangian $L$ may then be considered as
generalizations of the concept of geodesics to spinning space. In this
formulation the supplementary conditions then select those geodesics which
correspond to the world lines of the physical spinning particles.

The variation of the Lagrangian under arbitrary variations $(\del x^{\mu},
\del \ps^{a})$ is

\be
\ba{lll}
\del L & = & \dsp{ \del x^{\mu} \lh - g_{\mu\nu} \frac{D^{2}
x^{\nu}}{D\tau^{2}}
  - \frac{i}{2} \ps^{a}\ps^{b} R_{ab \mu\nu} \dot{x}^{\nu} \rh }\\
 & & \\
 & + & \dsp{ \Del \ps^{a}\, \eta_{ab}\, \frac{D\ps^{b}}{D\tau}\, +\,
\mbox{total
       derivative} } \\
\ea
\label{2.4}
\ee

\nit
For notational convenience we have introduced the covariantized variation of
$\ps^{a}$ \ct{RH1}

\be
\Del \ps^{a}\, =\, \del \ps^{a} - \del x^{\mu}\, \og_{\mu\:\:b}^{\:\:a}\,
                   \ps^{b}.
\label{2.5}
\ee

\nit
The equations of motion can be read off immediately from eq.(\ref{2.4}):

\be
\ba{rcl}
\dsp{  \frac{D^{2} x^{\mu}}{D\tau^{2}} } & = & \dsp{ \ddot{x}^{\mu} -
     \Gam_{\lb\nu}^{\:\:\:\:\:\mu}\, \dot{x}^{\lb} \dot{x}^{\nu}\, =\,
     - \frac{i}{2}\, \ps^{a}\ps^{b}\, R_{ab\:\:\,\nu}^{\:\:\:\:\,\mu}\,
     \dot{x}^{\nu} } \\
  &  &  \\
\dsp{ \frac{D\ps^{a}}{D\tau} } & = & 0.\\
\ea
\label{2.6}
\ee

\nit
The canonical momentum conjugate to $x^{\mu}$ is

\be
p_{\mu}\, =\, \dd{L}{\dot{x}^{\mu}}\, =\, g_{\mu\nu} \dot{x}^{\nu} -
   \frac{i}{2}\, \og_{\mu ab}\, \ps^{a} \ps^{b},
\label{2.7}
\ee

\nit
whilst the canonical momentum conjugate to $\ps^{a}$ is proportional to
$\ps^{a}$ itself:

\be
\pi_{a}\, =\, \dd{L}{\dot{\ps}^{a}}\, =\, - \frac{i}{2}\, \ps_{a}.
\label{2.8}
\ee

\nit
This implies a second-class constraint. Eliminating the constraint by
Dirac's procedure one obtains the canonical Poisson-Dirac brackets

\be
\left\{ x^{\mu}, p_{\nu} \right\}\, =\, \del^{\mu}_{\nu}, \hspace{3em}
\left\{ \ps^{a}, \ps^{b} \right\}\, =\, -i \eta^{ab}.
\label{2.9}
\ee

\nit
The Poisson-Dirac brackets for general functions $F$ and $G$ of the canonical
phase-space variables $(x, p, \ps)$ accordingly read

\be
\left\{ F, G \right\}\, =\, \dd{F}{x^{\mu}}\, \dd{G}{p_{\mu}}\, -\,
 \dd{F}{p_{\mu}}\, \dd{G}{x^{\mu}}\, +\, i\, (-1)^{a_{F}}\, \dd{F}{\ps^{a}}\,
 \dd{G}{\ps_{a}}.
\label{2.10}
\ee

\nit
Here $a_{F}$ is the Grassmann parity of $F$: $a_{F} = (0,1)$ for $F$ = (even,
odd). The canonical Hamiltonian of the theory is given by

\be
H\, =\, \frac{1}{2}\, g^{\mu\nu}\, \lh p_{\mu} + \og_{\mu} \rh \lh p_{\nu} +
   \og_{\nu} \rh,
\label{2.11}
\ee

\nit
where we define $\og_{\mu} = i/2\: \og_{\mu ab} \ps^{a} \ps^{b} $. Indeed, the
time-evolution of any function $F(x,p,\ps)$ is generated by its Poisson-Dirac
bracket with this Hamiltonian:

\be
\frac{d F}{d \tau}\, =\, \left\{ F, H \right\}.
\label{2.12}
\ee

\nit
Eqs.\ (\ref{2.9})-(\ref{2.12}) summarize the canonical structure of the theory.
In this formulation the fundamental brackets (\ref{2.9}) take their simplest
form. The disadvantage of the canonical formulation is, that one looses
manifest
covariance. For this reason it is often convenient to describe the theory in
terms of a set of covariant phase-space variables, defined by $x^{\mu}$,
$\ps^{a}$ and the covariant momentum

\be
\Pi_{\mu}\, \equiv\, p_{\mu}\, +\, \og_{\mu}\, =\, g_{\mu\nu}\,\dot{x}^{\nu}.
\label{2.13}
\ee

\nit
The Poisson-Dirac brackets for functions of the covariant phase-space
variables $(x,\Pi,\ps)$ then read

\be
\left\{ F, G \right\}\, =\,  \cD_{\mu} F\, \dd{G}{\Pi_{\mu}}\, -\,
  \dd{F}{\Pi_{\mu}}\, \cD_{\mu} G\, -\, R_{\mu\nu}\, \dd{F}{\Pi_{\mu}}\,
  \dd{G}{\Pi_{\nu}}\, +\, i\, (-1)^{a_{F}}\, \dd{F}{\ps^{a}}\, \dd{G}{\ps_{a}},
\label{2.14}
\ee

\nit
where the notation used is

\be
\ba{lll}
\dsp{ \cD_{\mu}\, F } & = & \dsp{ \partial_{\mu}\, F\, +\,
   \Gam_{\mu\nu}^{\:\:\:\:\:\lb}\, \Pi_{\lb}\, \dd{F}{\Pi_{\nu}}\, +\,
   \og_{\mu\:\:b}^{\:\:a}\, \ps^{b}\, \dd{F}{\ps^{a}}, } \\
  &  & \\
R_{\mu\nu} & = & \dsp{ \frac{i}{2}\, \ps^{a} \ps^{b} R_{ab \mu\nu}. }\\
\ea
\label{2.15}
\ee

\nit
Note that in particular

\be
\left\{ \Pi_{\mu}, \Pi_{\nu} \right\}\, =\, - R_{\mu\nu},
\label{2.16}
\ee

\nit
which is the classical analogue of the Ricci-identity (in the absence of
torsion). In terms of the new covariant phase-space variables the Hamiltonian
becomes

\be
H\, =\, \frac{1}{2}\, g^{\mu\nu}\, \Pi_{\mu} \Pi_{\nu}.
\label{2.17}
\ee

\nit
The dynamical equation (\ref{2.12}) remains of course unchanged. The
constraints
(\ref{2.1}) and (\ref{2.3.1}) become

\be
2\, H\, =\, g^{\mu\nu}\, \Pi_{\mu} \Pi_{\nu}\, =\, -1, \hspace{3em}
{\cal Q}\, =\, \Pi \cdot \ps\, =\,0.
\label{2.18}
\ee

\nit
Again these are to be imposed only after solving the theory, since they are
not compatible with the Poisson-Dirac brackets in general. However, one easily
establishes that

\be
\left\{ {\cal Q}, H \right\}\, =\, 0.
\label{2.19}
\ee

\nit
This establishes the conservation of ${\cal Q}$, whilst the Hamiltonian itself
is trivially conserved. Therefore the values of $H$ and $\cQ$ as chosen in
(\ref{2.18}) are preserved in time, and the physical conditions we impose are
consistent with the equations of motion\footnote{A more detailed discussion of
this point can be found in the second reference \ct{RH2}.}.

\section{Symmetries and Constants of Motion}{\label{S.3}}

The theory described by the Lagrangian (\ref{2.2}), or equivalently the
Hamiltonian (\ref{2.17}), admits a number of symmetries which are very useful
in
obtaining explicit solutions of the equations of motion \ct{RH3}, in particular
because of the connection with constants of motion via Noether's theorem. The
symmetries of a spinning-particle model can be divided into two classes:\nl
-- {\em generic} symmetries, which exist for any space-time metric
$g_{\mu\nu}(x)$; \nl
-- {\em non-generic} symmetries, which depend on the explicit form of the
metric. \nl
In refs.\ \ct{RH1,RH2} it was shown that the theory defined by the Lagrangian
(\ref{2.2}) possesses 4 generic symmetries: proper-time translations, generated
by the Hamiltonian $H$; supersymmetry generated by the supercharge ${\cal Q}$,
eq.(\ref{2.18}); and furthermore chiral symmetry, generated by the chiral
charge

\be
\Gam_{\star}\, =\, - \frac{ i^{[\frac{d}{2}]} }{d!}\, \ve_{a_{1}...a_{d}}\,
   \ps^{a_{1}}\, ... \ps^{a_{d}},
\label{3.1}
\ee

\nit
and dual supersymmetry, generated by the dual supercharge

\be
{\cal Q}^{\star}\, =\, i\, \left\{ {\cal Q}, \Gam_{\star} \right\}\, =\,
  \frac{ -i^{[\frac{d}{2}]} }{(d-1)!}\, \ve_{a_{1}...a_{d}}\, e^{\mu a_{1}}\,
  \Pi_{\mu}\, \ps^{a_{2}}\, ... \ps^{a_{d}}.
\label{3.2}
\ee

\nit
It is straightforward to check, that all these quantities have vanishing
Poisson-Dirac brackets with the Hamiltonian, and therefore are constants of
motion.

To obtain all symmetries, including the non-generic ones, one has to find all
functions ${\cal J}(x,\Pi,\ps)$ which commute with the Hamiltonian in the sense
of Poisson-Dirac brackets:

\be
\left\{ H, \cJ \right\}\, =\, 0.
\label{3.3}
\ee

\nit
The covariant form (\ref{2.14}) of the brackets immediately gives

\be
\Pi^{\mu}\, \left( \cD_{\mu}\, {\cal J}\, +\, R_{\mu\nu}
  \dd{{\cal J}}{\Pi_{\nu}} \right)\, =\, 0.
\label{3.4}
\ee

\nit
Note that if $\cJ$ depends on the covariant momentum only via the Hamiltonian:
$\cJ(x,\Pi,\ps) = \cJ(x,H,\ps)$, then the second term vanishes identically
and the equation simplifies to

\be
\Pi \cdot \cD\, \cJ\, =\, 0.
\label{3.4.1}
\ee

\nit
In all other cases we require the full eq.(\ref{3.4}). If we expand ${\cal J}$
in a power series in the covariant momentum

\be
{\cal J}\, =\, \sum_{n=0}^{\infty}\, \frac{1}{n!}\,
      J^{(n)\, \mu_{1}...\mu_{n}}(x,\ps)\, \Pi_{\mu_{1}}\, ... \Pi_{\mu_{n}},
\label{3.5}
\ee

\nit
then eq.(\ref{3.4}) is satisfied for arbitrary $\Pi_{\mu}$ if and only if the
components of ${\cal J}$ satisfy

\be
D_{\left( \mu_{n+1} \right.}\, J^{(n)}_{\left. \mu_{1}...\mu_{n} \right)}\,
   +\, \og_{\lh \mu_{n+1} \rd \:\:b}^{\:\:\:\:a}\, \ps^{b}\,
   \dd{J^{(n)}_{\ld \mu_{1} ... \mu_{n} \rh}}{\ps^{a}}
   =\, R_{\nu \left(\mu_{n+1}\right. }\, J^{(n+1)\:\:\:\:\nu}_{\left.
          \mu_{1}... \mu_{n} \right) },
\label{3.6}
\ee

\nit
where the parentheses denote full symmetrization over the indices enclosed.
Eqs.(\ref{3.6}) are the generalizations of the Killing equations to spinning
space first obtained in \ct{RH1}. An important aspect of these equations is,
that with the exception of the case described by eq.(\ref{3.4.1}), they couple
spinning-space Killing tensors $J^{(n)}$ of different rank,
unlike the case of Killing tensors in ordinary space.

We also observe, that any constant of motion ${\cal J}$ satisfies

\be
\ba{lll}
\dsp{ \left\{ \cQ, \cJ \right\} } & = & \dsp{ - \ps^{\mu}\, \left( \cD_{\mu}\,
  {\cal J}\, +\, R_{\mu\nu} \dd{{\cal J}}{\Pi_{\nu}} \right)\, -\, i\,
  e^{\mu a}\, \Pi_{\mu}\, \dd{{\cal J}}{\ps^{a}} } \\
  &  & \\
  & = & \dsp{ - \lh \ps \cdot \cD\, {\cal J}\, +\, i\, \Pi \cdot
              \dd{{\cal J}}{\ps} \rh, }\\
\ea
\label{3.7}
\ee

\nit
where the last line follows from the observation that the curvature term
contains three contractions with the anti-commuting spin variables, in
combination with the Bianchi identity $R_{\left[ \mu\nu\lb \right] \kg} = 0$.
Taking in particular ${\cal J} = {\cal Q}$ one finds

\be
\left\{ {\cal Q}, {\cal Q} \right\}\, =\, - 2 i H,
\label{3.8}
\ee

\nit
the conventional supersymmetry algebra. From this result and the
Jacobi-identity
for 2 ${\cal Q}$'s and any constant of motion ${\cal J}$ it follows, that

\be
\Th\, \equiv\, \left\{ \cQ, \cJ \right\}
\label{3.9}
\ee

\nit
is a superinvariant and hence a constant of motion as well:

\be
\left\{ \cQ, \Th \right\}\, =\, 0, \hspace{3em}
\left\{ H, \Th \right\}\, =\, 0.
\label{3.10}
\ee

\nit
Thus we infer that constants of motion generally come in supermultiplets
$({\cal J}, \Th)$, of which the prime example is the multiplet $({\cal Q}, H)$
itself. The only exceptions to this result are the constants of motion whose
bracket with the supercharge vanishes $(\Th = 0)$, but which are not themselves
obtained from the bracket of ${\cal Q}$ with another constant of motion.

   From eq.(\ref{3.7}) it follows that a superinvariant is a solution of the
equation

\be
\left\{ \cQ, \cJ \right\}\, =\, - \lh \ps \cdot \cD\, \cJ\, +\,
    i\, \Pi \cdot \dd{{\cal J}}{\ps} \rh\, =\, 0.
\label{3.11}
\ee

\nit
Let us write for ${\cal J}$ the series expansion

\be
{\cal J}(x,\Pi,\ps)\, =\, \sum_{m,n = 0}^{\infty}\, \frac{i^{\left[ \frac{m}{2}
   \right]}}{m!n!}\, \ps^{a_{1}}...\ps^{a_{m}}\, f^{(m,n)\,
\mu_{1}...\mu_{n}}_{
   \: a_{1}...a_{m} }(x)\, \Pi_{\mu_{1}}...\Pi_{\mu_{n}},
\label{3.11.1}
\ee

\nit
where $f^{(n,m)}$ is completely symmetric in the $n$ upper indices
$\{\mu_{k}\}$ and completely anti-symmetric in the $m$ lower indices
$\{a_{i}\}$; one then obtains the component equation

\be
n\, f^{(m+1,n-1)\, \lh \mu_{1}...\mu_{n-1} \right.}_{\: a_{0} a_{1}...a_{m}}\,
  e^{\left. \mu_{n} \right)\, a_{0} }\, =\, m\, D_{\left[ a_{1} \right.}\,
  f_{\left. \: a_{2}...a_{m} \right]}^{(m-1,n)\, \mu_{1}...\mu_{n}} ,
\label{3.11.2}
\ee

\nit
where $D_{a} = e^{\mu}_{\:\:a} D_{\mu}$ are ordinary covariant derivatives, and
indices in parentheses are to be symmetrized completely, whilst those in square
brackets are to be anti-symmetrized, all with unit weight. Note in particular
for $m = 0$:

\be
f_{a}^{(1,n)\, \lh \mu_{1}...\mu_{n} \right.}\, e^{\left. \mu_{n+1} \right)\,
a}
  \, =\, 0.
\label{3.11.2.1}
\ee

\nit
In a certain sense these equations represent a square root of the generalized
Killing equations, although they only provide sufficient, not necessary
conditions for obtaining solutions. However, of each supermultiplet (singlet
or non-singlet) at least {\em one} component is a solution of equation
(\ref{3.11}). Having found $\Th$ we can then try to reconstruct the
corresponding $\cJ$ by solving (\ref{3.9}).

The content of eqs.(\ref{3.11.2}) is twofold. On the one hand they partly solve
$f^{(m+1,n-1)}$ in terms of $f^{(m-1,n)}$. Only that part of $f^{(m+1,n-1)}$
is solved which is symmetrized in one flat index and all $(n-1)$ curved
indices. On the other hand eqs.(\ref{3.11.2}) do not automatically imply that
$f^{(m+1,n-1)}$ is completely anti-symmetric in the first $(m+1)$ indices.
Imposing that condition on eqs.(\ref{3.11.2}) one finds a new set of equations
which are precisely the generalised Killing equations for that part of
$f^{(m+1,n-1)}$ which was {\em not} given in terms of $f^{(m-1,n)}$, and which
should still be solved for. This is the part of $f^{(m+1,n-1)}$ which is
anti-symmetized in one curved index and all $(m+1)$ flat indices.

Hence eqs.(\ref{3.11.2}) clearly have advantages over the generalized Killing
equations (\ref{3.6}). In order to find the constant of motion corresponding to
a Killing tensor of rank $n$:

\be
\cD_{\lh \mu_{n+1} \rd}\, \cJ^{(n)}_{\ld \mu_{1} ... \mu_{n} \rh}\, =\, 0,
\label{3.11.2.2}
\ee

\nit
one has to solve the complicated hierarchy of partial differential equations
(\ref{3.6}) for $(\cJ^{(n-1)}, ... , \cJ^{(0)})$ and add the terms, as in
expression (\ref{3.5}). However, if we have a solution
$f_{a_{1}...a_{m}}^{(m,n)\, \mu_{1}...\mu_{n} }$ of the equation

\be
f_{a_{1}...a_{m}}^{(m,n)\, \lh \mu_{1}...\mu_{n} \rd}\, e^{\ld \mu_{n+1} \rh
   a_{1}}\, =\, 0,
\label{3.11.2.3}
\ee

\nit
then we generate at least part of the components
$f_{a_{1}...a_{m+2\ag}}^{(m+2\ag, n-\ag)\,\mu_{1} ... \mu_{n-\ag}}$ for
$\ag = 1, ..., n$ by mere differentiation. Eq.(\ref{3.11.1}) then gives us the
corresponding constant of motion. In section \ref{S.5} we consider an example
in which these advantages become clear.

Finally we observe, that eqs.(\ref{3.9}),(\ref{3.10}) imply that the
Poisson-Dirac brac\-ket with ${\cal Q}$ defines a nilpotent operation in the
space of constants of motion. Hence the supersinglets span the cohomology of
the
supercharge, whilst the supermultiplets $({\cal J}, \Th)$ form pairs of ${\cal
Q}$-exact and ${\cal Q}$-coexact forms. The solutions of eq.(\ref{3.11}) then
correspond to the $\cQ$-closed forms.

\section{New Supersymmetries}{\label{S.4}}

The constants of motion generate infinitesimal transformations of the
co-ordinates leaving the equations of motion invariant:

\be
\del x^{\mu}\, =\, \del \ag\, \left\{ x^{\mu}, \cJ \right\}, \hspace{3em}
\del \ps^{a}\, =\, \del \ag\, \left\{ \ps^{a}, \cJ \right\},
\label{4.1}
\ee

\nit
with $\del \ag$ the infinitesimal parameter of the transformation. In
particular, the action as defined by $L$, eq.(\ref{2.2}), is invariant under
the generic symmetries, such as supersymmetry:

\be
\ba{lll}
\del x^{\mu} & = & \dsp{ i \eps \left\{ \cQ, x^{\mu} \right\}\, =\, -i \eps\,
   e^{\mu}_{\:\:a}\, \ps^{a}, } \\
  &  &  \\
\del \ps^{a} & = & \dsp{ i \eps \left\{ \cQ, \ps^{a} \right\}\, =\, \eps\,
   e_{\mu}^{\:\:a}\, \dot{x}^{\mu}\, +\, \del x^{\mu}\, \og_{\mu\:\:b}^{\:\:a}
   \ps^{b}, }\\
\ea
\label{4.2}
\ee

\nit
where the infinitesimal parameter $\eps$ of the transformation is
Grassmann-odd.

We now ask whether the theory might admit other (non-generic) supersymmetries
of the type

\be
\del x^{\mu}\, =\, -i \eps\, f^{\mu}_{\:\:a}\, \ps^{a}\,
               \equiv\, - i \eps\, J^{(1)\mu}.
\label{4.3}
\ee

\nit
Such a transformation is generated by a phase-space function ${\cal Q}_{f}$

\be
{\cal Q}_{f}\, =\, J^{(1) \mu}\, \Pi_{\mu}\, +\, J^{(0)},
\label{4.4}
\ee

\nit
where $J^{(0,1)}(x,\ps)$ are independent of $\Pi$. Inserting
this Ansatz into the generalized Killing equations (\ref{3.6}), one finds

\be
J^{(0)}(x,\ps)\, =\, \frac{i}{3!}\, c_{abc}(x)\, \ps^{a} \ps^{b} \ps^{c},
\label{4.5}
\ee

\nit
with the tensors $f^{\mu}_{\:\:a}$ and $c_{abc}$ satisfying the conditions

\be
D_{\mu}\, f_{\nu a}\, +\, D_{\nu}\, f_{\mu a}\, =\, 0,
\label{4.6}
\ee

\nit
and

\be
D_{\mu}\, c_{abc}\, =\, - \lh R_{\mu\nu ab}\, f^{\nu}_{\:\:c}\, +\,
   R_{\mu\nu bc}\, f^{\nu}_{\:\:a}\, +\, R_{\mu\nu ca}\, f^{\nu}_{\:\:b}\, \rh
{}.
\label{4.7}
\ee

\nit
If we have $N$ symmetries of this kind, specified by $N$ sets of tensors
$( f_{i\: a}^{\mu}, c_{i\, abc})$, $i = 1,...,N$, then the corresponding
generators are

\be
\cQ_{i}\, =\, f_{i\, a}^{\mu}\, \Pi_{\mu}\, \ps^{a}\, +\, \frac{i}{3!}\,
      c_{i\, abc}\, \ps^{a} \ps^{b} \ps^{c}.
\label{4.8}
\ee

\nit
Observe, that the supercharge (\ref{2.18}) is precisely of this form
with $f^{\mu}_{\:\:a} = e^{\mu}_{\:\:a}$ and $c_{abc} = 0$. For notational
convenience we sometimes refer to these quantities defining the standard
supersymmetry by assigning them the index $i = 0$: $\cQ = \cQ_{0}$,
$e^{\mu}_{\:\:a} = f^{\mu}_{0\, a}$, etc.

   From the general result (\ref{2.14}) one now obtains the following algebra
of
Poisson-Dirac brackets of the conserved charges $\cQ_{i}$:

\be
\left\{ {\cal Q}_{i}, {\cal Q}_{j} \right\}\, =\, -\, 2i\, Z_{ij},
\label{4.9}
\ee

\nit
with

\be
Z_{ij}\, =\, \frac{1}{2}\, K_{ij}^{\mu\nu}\, \Pi_{\mu} \Pi_{\nu}\, +\,
             I^{\mu}_{ij}\, \Pi_{\mu}\, +\, G_{ij},
\label{4.10}
\ee

\nit
and

\be
\ba{lll}
K^{\mu\nu}_{ij}  & = & \dsp{ \frac{1}{2}\, \lh f_{i\, a}^{\mu} f_{j}^{\nu a}
                       + f_{i\, a}^{\nu} f_{j}^{\mu a} \rh, } \\
  &  &  \\
I^{\mu}_{ij} & = & \dsp{ \frac{i}{2}\, \ps^{a} \ps^{b}\, I^{\mu}_{ij\, ab} }\\
  &  &  \\
  & = & \dsp{ \frac{i}{2}\, \ps^{a} \ps^{b}\, \lh f^{\nu}_{i\, b} D_{\nu}
        f_{j\, a}^{\mu} + f^{\nu}_{j\, b} D_{\nu} f_{i\, a}^{\mu} +
        \frac{1}{2}\, f_{i}^{\mu c} c_{j\, abc} + \frac{1}{2}\, f_{j}^{\mu c}
        c_{i\, abc} \rh, } \\
  &  &  \\
G_{ij} & = & \dsp{ - \frac{1}{4}\, \ps^{a} \ps^{b} \ps^{c} \ps^{d}\,
             G_{ij\, abcd} }\\
  &  &  \\
  & = & \dsp{ - \frac{1}{4}\, \ps^{a} \ps^{b} \ps^{c} \ps^{d}\,
        \lh R_{\mu\nu ab} f^{\mu}_{i\, c} f^{\nu}_{j\, d} + \frac{1}{\,2}
        c_{i\, ab}^{\:\:\:\:\:\: e} c_{j\, cde} \rh. }\\
\ea
\label{4.11}
\ee

\nit
An explicit calculation shows, that $K_{ij\, \mu\nu}$ is a symmetric Killing
tensor of 2nd rank :

\be
D_{\lh \lb \rd}\, K_{ij\, (\ld \mu\nu \rh} \, =\, 0,
\label{4.12}
\ee

\nit
whilst $I^{\mu}_{ij}$ is the corresponding Killing vector:

\be
\cD_{\lh \mu \rd}\, I_{ij\, \ld \nu \rh}\, =\,  \frac{i}{2}\, \ps^{a} \ps^{b}\,
    D_{\left( \mu \right.}\, I_{ij\, \left. \nu \right) ab}\, =\, \frac{i}{2}\,
    \ps^{a} \ps^{b}\, R_{ab\lb \left( \mu \right.}\,
    K_{ij\, \ld \nu \rh}^{\:\:\:\:\:\: \lb},
\label{4.13}
\ee

\nit
and $G_{ij}$ the corresponding Killing scalar:

\be
\cD_{\mu}\, G_{ij}\, =\, - \frac{1}{4}\, \ps^{a} \ps^{b} \ps^{c} \ps^{d}\,
  D_{\mu}\, G_{ij\, abcd}\, =\, \frac{i}{2}\, \ps^{a} \ps^{b}\,
  R_{ab\lb \mu}\, I^{\lb}_{ij}.
\label{4.14}
\ee

\nit
Since the Grassmann-even phase-space functions $Z_{ij}$ satisfy the generalized
Killing equations, their bracket with the Hamiltonian vanishes and they are
constants of motion:

\be
\frac{d Z_{ij}}{d \tau}\, =\, 0.
\label{4.15}
\ee

\nit
Note that for $i = j = 0$ we reobtain the usual supersymmetry algebra:

\be
\left\{ {\cal Q}, {\cal Q} \right\}\, =\, -\, 2i\, H,
\label{4.16}
\ee

\nit
where $H$ is the Hamiltonian. The $Z_{ij}$ with $i$ or $j$ not equal to zero
correspond to new bosonic symmetries, unless $K_{ij}^{\mu\nu} = \lb_{(ij)}\,
g^{\mu\nu}$, with $\lb_{(ij)}$ a constant (maybe zero). In that case the
corresponding Killing vector $I^{\mu}_{ij}$ and Killing scalar $G_{ij}$ vanish
identically; moreover if $\lb_{(ij)} \neq 0$, the corresponding supercharges
close on the Hamiltonian, which proves the existence of a second supersymmetry
of the standard type; we then have an $N$-extended supersymmetry, with $N \geq
2$. On the other hand, if there exists a second independent Killing tensor
$K^{\mu\nu}$ not proportional to $g^{\mu\nu}$, then we obtain a genuine new
type
of supersymmetry.

Now, as proven in sect.\ {\ref{S.4}}, the bracket of a supersymmetry $\cQ_{i}$
with the original supercharge $\cQ$ vanishes, and hence $\cQ_{i}$ is
superinvariant, if and only if

\be
K_{0i}^{\mu\nu}\, =\, f^{\mu}_{\:\:a}\, e^{\nu a}\, +\, f^{\nu}_{\:\:a}\,
  e^{\mu a}\, =\, 0.
\label{4.16.1}
\ee

\nit
In the language of $\cQ$-cohomology, $\cQ_{i}$ is $\cQ$-closed; according to
the
discussion at the end of sect.\ {\ref{S.2}} we can then construct the full
constant of motion $Z_{ij}$ directly by repeated differentiation of
$f_{\:\:a}^{\mu}$.

Finally, since the $Z_{ij}$ are symmetric in $(ij)$ by construction we can
diagonalize them and thus obtain an algebra\footnote{No summation over repeated
indices is implied on the right-hand side.}

\be
\left\{ {\cal Q}_{i}, {\cal Q}_{j} \right\}\, =\, -\, 2i\,\del_{ij}\, Z_{i},
\label{4.17}
\ee

\nit
with $N + 1$ conserved bosonic charges $Z_{i}$. If condition (\ref{4.16.1}) is
satisfied for all $\cQ_{i}$, the first of these diagonal charges (with $i = 0$)
is the Hamiltonian: $Z_{0} = H$.

\section{Properties of the $f$-symbols}{\label{S.5}}

In order to study the properties of the new supersymmetries, we now turn our
attention to the quantities $f^{\mu}_{\:\:a}$. It is convenient to introduce
the
2nd rank tensor

\be
f_{\mu\nu}\, =\, f_{\mu a}\, e^{\:\:a}_{\nu},
\label{5.0}
\ee

\nit
which will be refered to as the $f$-symbol. The defining relation (\ref{4.6})
implies

\be
D_{\nu}\, f_{\lb \mu}\, +\, D_{\lb}\, f_{\nu \mu}\, =\, 0.
\label{5.1}
\ee

\nit
It follows that the $f$-symbol is divergence-less on its first index

\be
D_{\nu}\, f^{\nu}_{\:\:\mu}\,=\, 0.
\label{5.2}
\ee

\nit
By contracting of eq.(\ref{5.1}) one finds

\be
D_{\nu}\, f_{\mu}^{\:\:\nu}\, =\, - \partial_{\mu}\, f_{\nu}^{\:\:\nu}.
\label{5.3}
\ee

\nit
Hence the divergence on the second index vanishes if and only if the trace of
the $f$-symbol is constant:

\be
D_{\nu}\, f_{\mu}^{\:\:\nu}\, =\, 0 \hspace{1em} \Leftrightarrow \hspace{1em}
          f_{\mu}^{\:\:\mu}\, =\, const.
\label{5.4}
\ee

\nit
Now observe, that the metric tensor $g_{\mu\nu}$ is a trivial solution of
eq.(\ref{5.1}); therefore if the trace is constant, it maybe subtracted from
the $f$-symbol without spoiling condition (\ref{5.1}). It follows, that in this
case one may without loss of generality always take the constant equal to zero
and hence $f$ to be traceless.

The symmetric part of the $i$th $f$-symbol is the tensor

\be
S_{\mu\nu}\, \equiv\, K_{i0\, \mu\nu}\, =\, \frac{1}{2}\, \lh f_{\mu\nu} +
   f_{\nu\mu} \rh,
\label{5.4.1}
\ee

\nit
defined in the first eq.(\ref{4.11}) with $f^{\mu}_{0\, a} = e^{\mu}_{\:\:a}$.
As was discussed there, it satisfies the generalized Killing equation

\be
D_{\left( \mu \right.}\, S_{\left. \nu \lb \right)}\, =\, 0.
\label{5.6}
\ee

\nit
We can also construct the anti-symmetric part

\be
B_{\mu\nu}\, =\, - B_{\nu\mu}\,
             =\, \frac{1}{2}\, \lh f_{\mu\nu} - f_{\nu\mu} \rh.
\label{5.7}
\ee

\nit
It obeys the condition

\be
D_{\nu}\, B_{\lb\mu}\, +\, D_{\lb}\, B_{\nu\mu}\, =\, D_{\mu}\, S_{\nu\lb}.
\label{5.8}
\ee

\nit
It follows, that if the symmetric part does not vanish and is not covariantly
constant, then the anti-symmetric part $B_{\mu\nu}$ by itself is {\em not} a
solution of eq.(\ref{5.1}). But by the same token the anti-symmetric part of
$f$ can not vanish either, hence $f$ can be completely symmetric only if it is
covariantly constant. \nl

It is of considerable interest to study the case in which the $f$-symbol is
completely anti-symmetric: \ $f_{\mu\nu}\, =\, B_{\mu\nu}$. This is precisely
the condition (\ref{4.16.1}) for the supercharge $\cQ_{f}$ to anti-commute with
ordinary supersymmetry in the sense of Poisson-Dirac brackets. In this case
also eq.(\ref{5.4}) is satisfied automatically.

For anti-symmetric $f_{\mu\nu}$ it is possible to say much more about the
explicit form of the quantities that were introduced above. First of all, if
the symmetric part of a certain $f_{i\:\mu\nu}$ vanishes:

\be
S_{i}^{\mu\nu}\, =\, K_{i0}^{\mu\nu} = 0,
\label{5.9}
\ee

\nit
then the corresponding Killing vector $I_{i0}^{\mu}$ and the Killing scalar
$G_{i0}$ vanish as well. Hence for this particular value of $i$ the complete
$Z_{i0} = 0$, and therefore

\be
\left\{ {\cal Q}_{i}, {\cal Q} \right\}\, =\, 0,
\label{5.10}
\ee

\nit
showing that $\cQ_{i}$ is superinvariant. To prove these assertions, we first
note that for anti-symmetric $f^{\mu\nu}$ eq.(\ref{5.1}) becomes

\be
D_{\nu}\, B_{\lb \mu}\, =\, - D_{\lb}\, B_{\nu \mu}.
\label{5.12}
\ee

\nit
Together with the anti-symmetry of $B_{\mu\nu}$ it follows that the gradient
is completely anti-symmetric:

\be
D_{\mu}\, B_{\nu\lb}\, =\, D_{\left[ \mu \right.}\, B_{\ld \nu \lb \right]}\,
                       \equiv H_{\mu\nu\lb}.
\label{5.13}
\ee

\nit
By taking the second covariant derivative of $f_{\mu\nu}$, commuting the
derivatives and using eq.(\ref{5.1}) we derive the identity

\be
D_{\mu}\, D_{\nu}\, f_{\lb\kg}\, =\, R_{\nu\lb\mu}^{\:\:\:\:\:\:\:\sg}\,
  f_{\sg\kg}\, +\, \frac{1}{2}\, \lh R_{\nu\lb\kg}^{\:\:\:\:\:\:\:\sg}\,
  f_{\mu\sg}\, +\, R_{\mu\lb\kg}^{\:\:\:\:\:\:\:\sg}\, f_{\nu\sg}\, -\,
  R_{\mu\nu\kg}^{\:\:\:\:\:\:\:\sg}\, f_{\lb\sg} \rh.
\label{5.14}
\ee

\nit
For the special case of anti-symmetric $f_{\mu\nu}$ this implies

\be
D_{\mu}\, H_{\nu\lb\kg}\, =\, \frac{1}{2}\, \lh
  R_{\nu\lb\mu}^{\:\:\:\:\:\:\:\sg}\, f_{\sg\kg}\, +\,
  R_{\lb\kg\mu}^{\:\:\:\:\:\:\:\sg}\, f_{\sg\nu}\, +\,
  R_{\kg\nu\mu}^{\:\:\:\:\:\:\:\sg}\, f_{\sg\lb} \rh.
\label{5.15}
\ee

\nit
Comparison with eq.(\ref{4.7}) shows, that

\be
-\frac{1}{2}\, c_{abc}\, =\, H_{abc}\, =\,
   e^{\mu}_{\:\:a} e^{\nu}_{\:\:b} e^{\lb}_{\:\:c}\, H_{\mu\nu\lb},
\label{5.16}
\ee

\nit
modulo a covariantly constant term. This result is an instance of
eq.(\ref{3.11.2}) with $n = 1$, $m = 2$. Note that we can always choose the
covariantly constant term to vanish, since in order to construct a constant
of motion we only need a particular solution of eq.(\ref{4.7}). \nl

As a side-remark we note, that if a covariantly constant three-index tensor
$c_{abc}$ exists, then it always provides us with another symmetry,
corresponding to the Killing vector

\be
I_{\mu}\, =\, \frac{i}{2}\, \ps^{a} \ps^{b}\, e_{\mu}^{\:\: c}\, c_{abc}.
\label{5.18}
\ee

\nit
More precisely, if $D_{\mu}\, c_{abc}\, =\, 0$ then

\be
\cD_{\mu}\, I_{\nu}\, =\, 0,
\label{5.19}
\ee

\nit
and the generalized Killing equation is automatically satisfied for $I_{\mu}$.
In this case we are free to add the term with $c_{abc}$ to the supercharge, but
it is not required, since both terms are conserved separately. \nl

Returning to eq.(\ref{5.10}), we first observe that according to eq.(\ref{5.9})
$K_{0i}^{\mu\nu} = 0$. Moreover, since $c_{0\, abc} = 0$ identically, the
r.h.s.\ of the second of eqs.(\ref{4.11}) now becomes

\be
I_{i0\,\mu\nu\lb}\, \equiv\, I_{i0\,\mu ab}\, e^{\:\:a}_{\nu}\,
e^{\:\:b}_{\lb}\,
  =\, D_{\lb}\, B_{i\, \mu\nu}\, +\, \frac{1}{2}\, c_{i\, \mu\nu\lb}\, =\, 0.
\label{5.17}
\ee

\nit
The last equality follows from eq.(\ref{5.16}). Finally, the Killing scalar
$G_{i0}$ vanishes because of the cyclic Bianchi identity for the Riemann tensor
$R_{\mu\nu\lb\kg}$ and the vanishing of at least one of the three-index
tensors:
$c_{0\, abc} = 0$. This proves eq.(\ref{5.10}).

Anti-symmetric $f$-symbols and their corresponding Killing-tensors have been
studied extensively in refs.\ct{Carter,CMcL} in the related context of finding
solutions of the Dirac-equation in non-trivial curved space-time. They
correspond precisely to the Killing-Yano and Stackel-Killing tensors described
in these papers. The analysis presented here shows, that they belong to a
larger class of possible structures which generate generalized supersymmetry
algebras.

\section{The Kerr-Newman metric}{\label{S.8}}

In this section we apply the results obtained previously to show that a new
kind of supersymmetry exists in spinning Kerr-Newman space.

The gravitational and electromagnetic field of a rotating particle with mass
$M$ and charge $Q$ are described by the Kerr-Newmann metric, which reads

\be
\ba{lll}
\dsp{ds^2} & = &
\dsp{ - \, \frac{\Del}{\rg^2} \left[ dt - a \sin^2 \th \, d \vf \right]^2
+ \frac{\sin^2 \th}{\rg^2} \left[ (r^2 + a^2) d \vf - a dt \right]^2 + \nb} \\
 & & \\
 & & \dsp{ + \, \frac{\rg^2}{\Del} dr^2 + \rg^2 d\th^2 ,}
\ea
\label{8.1}
\ee

\nit
and the electromagnetic field tensor

\be
\ba{lll}
\dsp{F} & = & \dsp{\frac{Q}{\rg^4} (r^2 - a^2 \cos^2 \th) dr \wedge
\left[ dt - a \sin^2 \th \, d \vf \right] + \nb }\\
 & & \\
 & + &  \dsp{\frac{2Qar \cos \th \sin \th}{\rg^4} d\th \wedge
\left[ - a dt + (r^2 + a^2) d\vf  \right] .}
\label{8.2}
\ea
\ee

\nit
Here

\be
\ba{lll}
\dsp{\Del} & = & \dsp{r^2 + a^2 - 2Mr + Q^2 , \nb }\\
 & & \\
\dsp{\rg^2} & = & \dsp{r^2 + a^2 \cos^2 \th ,}
\ea
\label{8.3}
\ee

\nit
and the total angular momentum is $J = Ma$. The expressions for $ds^2$ and $F$
only describe the fields {\em outside} the horizon, which is located at

\be
r = M + \sqrt{M^2 - Q^2 - a^2} .
\label{8.4}
\ee

\nit
As was found by Carter \ct{Carter2}, the Kerr-Newman space has two independent
second rank Killing tensors. The metric tensor $g_{\mu \nu}$, here defined by
eq.(\ref{8.1}), is a Stackel-Killing tensor for any geometry and the
corresponding conserved quantity is the Hamiltonian $H$. Furthermore there
exists
another Stackel-Killing tensor $K_{\mu \nu}$, which corresponds to a conserved
quantity $Z$. A supersymmetric extension of this result, applying to spinning
particles, is based on the anti-symmetric Killing-Yano tensor $f_{\mu \nu}$
found by Penrose and Floyd \ct{PF}, which satisfies eq.(\ref{5.1}):

\[
D_{\lb}\, f_{\mu \nu}\, +\, D_{\mu}\, f_{\lb \nu}\, =\, 0,
\]

\nit
and the covariant square of which is exactly the Stackel-Killing tensor
$K_{\mu \nu}$. The new supersymmetry in spinning Kerr-Newmann space is then
generated by a supercharge of the form given in eq.(\ref{4.8}), with the
Killing-Yano tensor as the $f$-symbol of the double vector $f_{\mu}^{\:\:a}$

\[
f_{\mu}^{\:\:a} \, = \, f_{\mu \nu} e^{\nu a} ,
\]

\nit
and a corresponding three-index tensor $c_{abc}$ obtained as in
eq.(\ref{5.16}).

We now calculate the explicit expression for the new supercharge and use this
to find the Killing vector $I_{\mu}$ and the Killing scalar $G$ which
correspond
to the Stackel-Killing tensor $K_{\mu \nu}$ in spinning Kerr-Newman space and
which define the corresponding conserved charge $Z$.

The Killing-Yano tensor is defined by \ct{PF}

\be
\ba{lll}
\dsp{\frac{1}{2} \,  f_{\mu \nu} \, dx^{\mu} \wedge dx^{\nu}} & = &
\dsp{a \cos \th \, dr \wedge \left[ dt - a \sin^{2} \th \, d \vf \right] +} \\
 & & \\
& + & \dsp{r \sin \th \, d \th \wedge
\left[ - a dt + (r^{2} + a^{2}) d\vf \right].}
\ea
\label{8.7}
\ee

\nit
Using the expressions for the vielbein $e_{\mu}^{\;\;a}(x)$ corresponding to
the metric given in eq.(\ref{8.1})

\be
\ba{lll}
\dsp{e_{\mu}^{\;\;0} \, dx^{\mu} } & = & \dsp{ - \frac{\sqrt{\Del}}{\rg}
                               \left[dt - a \sin^{2} \th \, d \vf \right] , }
\\
 & & \\
\dsp{e_{\mu}^{\;\;1} \, dx^{\mu} } & = & \dsp{ \frac{\rg}{\sqrt{\Del}} dr ,} \\
 & & \\
\dsp{e_{\mu}^{\;\;2} \, dx^{\mu} } & = & \dsp{ \rg d \th , } \\
 & & \\
\dsp{e_{\mu}^{\;\;3} \, dx^{\mu} } & = & \dsp{\frac{\sin \th}{\rg}
\left[-a dt + (r^{2} + a^{2}) d\vf \right] ,}
\ea
\label{8.8}
\ee

\nit
one finds the following components of $f_{\mu}^{\;\;a}(x)$

\be
\ba{lll}
\dsp{f_{\mu}^{\;\;0} \, dx^{\mu} } & = &  \dsp{ \frac{\rg}{\sqrt{\Del}}
                                                      \,  a \cos \th dr ,} \\
 & & \\
\dsp{f_{\mu}^{\;\;1} \, dx^{\mu} } & = &  \dsp{ - \frac{\sqrt{\Del}}{\rg}
                \,  a \cos \th \left[dt - a \sin^{2} \th \, d \vf \right] , }\\
 & & \\
\dsp{f_{\mu}^{\;\;2} \, dx^{\mu} } & = & \dsp{ - \frac{r \sin \th}{\rg}
                            \left[-a dt + (r^{2} + a^{2}) d\vf \right] ,} \\
 & & \\
\dsp{f_{\mu}^{\;\;3} \, dx^{\mu} } & = & \rg r d \th
\ea
\label{8.9}
\ee

\nit
One can check that this $f_{\mu}^{\;\;a}(x)$ indeed satisfies eq.(\ref{4.6}).
Finally, to find a conserved quantity we need to calculate $c_{a b c}(x)$ from
eq.(\ref{5.16}). Its components read

\be
\ba{lllllll}
\dsp{c_{0 1 2}} & = & \dsp{\frac{2 a \sin \th}{\rg}} & \hspace{2cm}
 & \dsp{c_{0 1 3}} & = & 0 \\
 & & & & & & \\
\dsp{c_{0 2 3}} & = & 0 &
 & \dsp{c_{1 2 3}} & = & \dsp{- \frac{2 \sqrt{\Del}}{\rg}}
\ea
\label{8.10}
\ee

\nit
Inserting the quantities given in eqs.(\ref{8.9}),(\ref{8.10}) into
eq.(\ref{4.8}) we obtain the new supersymmetry generator $\cQ_{f}$ for spinning
Kerr-Newman space. From this expression and using eqs.(\ref{4.11}) we can
construct the Killing tensor, vector and scalar which define the conserved
charge $Z = i/2\, \{ \cQ_{f} , \cQ_{f} \}$. The results are

\begin{eqnarray}
\dsp{ K_{\mu \nu}(x) \, dx^{\mu} \, dx^{\nu}} & = &
      \dsp{ - \frac{\rg^{2} a^{2} \cos^{2} \th}{\Del} dr^{2}
      + \frac{\Del a^{2} \cos^{2} \th}{\rg^{2}}
      \left[dt - a \sin^{2} \th \, d \vf \right]^{2} + } \nonumber \\
 & & \nonumber \\
 & & \dsp{ + \frac{r^{2} \sin^{2} \th}{\rg^{2}}
        \left[-a dt + (r^{2} + a^{2}) d \vf \right]^{2}
      + \rg^{2} r^{2} d \th^{2},}
\label{8.11.1}
\\
 & & \nonumber \\
\dsp{ I_{\mu}(x) \, dx^{\mu} } & = & \dsp{   \frac{2i}{\rg^{2}}
        \left[ r \sin \th \, \psi^{1} + \sqrt{\Del} \cos \th \, \psi^{2}
\right]
        \left[ a \sin \th \, \psi^{0} - \sqrt{\Del} \, \psi^{3} \right]
                   \times } \nonumber\\
 & & \nonumber \\
 & & \dsp{ \hspace{2cm}
        \times \left[- a dt + (r^{2} + a^{2}) d \vf \right] + } \nonumber \\
 & & \nonumber \\
 & & \dsp{ - i \sqrt{\Del} \cos \th \, \psi^{2} \left[ a \sin \th \, \psi^{0} -
             \sqrt{\Del} \, \psi^{3} \right] d \vf } \nonumber \\
 & & \nonumber \\
 & & \dsp{ + i \sqrt{\Del}
        \left[ r \sin \th \, \psi^{1} + \sqrt{\Del} \cos \th \, \psi^{2}
\right]
        \psi^{3} d \vf + } \nonumber \\
 & & \nonumber \\
 & & \dsp{ + \frac{i a \sin \th}{\sqrt{\Del}} \left[ r \, \psi^{0} \psi^{3} +
           a \cos \th \, \psi^{1} \psi{2} \right] dr } \nonumber \\
 & & \nonumber \\
 & & \dsp{ + i \sqrt{\Del} \left[ a \cos \th \, \psi^{0} \psi^{3} - r \,
           \psi^{1} \psi^{2} \right] d \th, }
\label{8.11.2} \\
 & & \nonumber \\
\dsp{ G } & = & \dsp{ - \frac{2 Q a \cos \th}{\rg^{2}}
                \, \psi^{0} \psi^{1} \psi^{2} \psi^{3}. }
\label{8.11.3}
\end{eqnarray}

\nit
The physical interpretation of these equations may become more clear if we
recall, that the anti-commuting spin variables are related to the standard
anti-symmetric spin tensor $S^{ab}$, which appears in the definition of the
generators of the local Lorentz transformations, by

\be
S^{ab}\, =\, -i\, \ps^{a} \ps^{b}.
\label{8.12}
\ee

\nit
Indeed, from the Dirac-Poisson brackets (\ref{2.10}) it can be verified
straightforwardly that they satisfy the SO(3,1) algebra. The full Lorentz
transformations are then generated by $M^{ab} = L^{ab} + S^{ab}$, $L^{ab}$
being
the orbital part. Like the generators of the Lorentz algebra, the generators of
other symmetries like $Z$ now also receive spin-dependent contributions. The
Killing tensor $K_{\mu \nu}$ given in (\ref{8.11.1}) is the one which was found
in \ct{C}. For spinless particles in Kerr-Newman space it defines a constant of
motion directly, whereas for spinning particles it now requires the non-trivial
contributions from spin which involve the Killing vector and Killing scalar
computed above.

\np
\nit
{\bf Acknowledgement}\nl

\nit
The initial work for this paper was done while G.W.\ Gibbons was Kramers
visiting professor of theoretical physics at the University of Utrecht. He
would like to thank the members of the institute, and in particular prof.\
G.\ 't Hooft, for their hospitality during his stay. He would also like to
thank
Brandon Carter for many stimulating discussions over the years on the subject
of Killing tensors and John Stewart for help with algebraic computing.

\end{document}